\documentclass[12pt]{iopart}
\pdfoutput=1
\usepackage{multirow}
\usepackage{graphicx}
\usepackage{hyperref}
\usepackage{graphicx}
\usepackage{amssymb}
\usepackage{color}
\usepackage{cite}
\begin{document}

\title[Nucleosome remodeling motor]{A molecule that detects the length of DNA by using chain fluctuations}

\author{Kuni H Iwasa$^{1,2}$ and Ana Maria Florescu$^{1,3}$}
\address{$^1$ Max Planck Institute for Physics of Complex Systems,  N\"othnitzer Str.\ 38, Dresden, D-01187 Germany}
\address{$^2$ NIH Porter Neuroscience Research Center, 35A Convent Drive, Bethesda, Maryland 20892, USA
\ead{iwasa@mailaps.org}}
\address{$^3$ SISSA,  Via Bonomea 265, 34136 Trieste, Italy \ead{aflorescu@sissa.it}}

\begin{abstract}
A class of nucleosome remodeling motors translocate nucleosomes, to which they are attached, toward the middle of DNA chain in the presence of ATP during \emph{in vitro} experiments. Such a biological activity is likely based on a physical mechanism for detecting and comparing the lengths of the flanking polymer chains. Here we propose that a pivoting mode of DNA fluctuations near the surface of the nucleosome coupled with binding reaction with a DNA binding site of the motor provides a physical basis for length detection. Since the mean frequency of fluctuations is higher for a shorter chain than a longer one due to its lower drag coefficient, a shorter chain has a higher rate of receptor binding, which triggers the ATP-dependent activity of the remodeling motor. Dimerization of such units allows the motor to compare the length of the flanking DNA chains, enabling the translocation of the nucleosome toward the center of the DNA.
\end{abstract}

\maketitle


\section{INTRODUCTION}
Chromatin remodeling motors translocate nucleosomes, which are the first level of DNA packaging into chromatin in eukaryotic cells, along the associated DNA \cite{Clapier2009}.  The core of each nucleosome is a positively charged histone octamer, which is tightly wrapped around by 147 base pairs (bp) of negatively charged DNA, rendering those base pairs inaccessible to proteins for transcription or repair. For this reason, the positioning of nucleosomes constitutes a part of cellular memory \cite{Hwang2014}. Remodeling motors, which uses ATP as the energy source, enables various cellular functions by making these base pairs accessible by translocating the nucleosomes along the DNA \cite{Leschziner2011}. 

\begin{figure}[h]
\begin{center}
\begin{minipage}[b]{4cm}\textbf{A} \end{minipage}\\
\includegraphics[width=3.8cm]{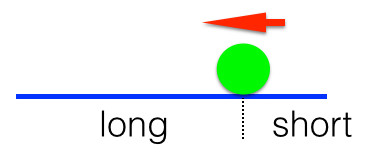}\\ \vspace{3mm}
\begin{minipage}[b]{4cm}\textbf{B} \end{minipage}\\ 
\includegraphics[width=3.8cm]{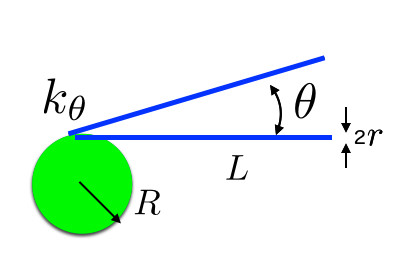}\\  
\label{fig:fluc}
\caption{Movement of nucleosome on DNA associated with the remodeling motor and pivoting motion of DNA chain on a nucleosome as a putative sensing mechanism of chain length. A: The remodeling motor moves its associated nucleosome  (circle) along the DNA (solid line) to the center. B:  Motion of DNA (lines) of length $L$ pivoting at the surface of a nucleosome (circle) of radius $R$. The spring constant of the pivot is $k_\theta$.}
\end{center}
\end{figure}

Interestingly, a class of remodeling motors known as the ISWI family simply arranges nucleosomes equidistantly \cite{Clapier2009,Leschziner2011}, somewhat reminiscent of wiping out memory from a hard disk of computers.  That biological function appears to be associated with translocating nucleosomes in the direction that the chain length of the flanking DNA is longer (Fig.\ \ref{fig:fluc}A). Indeed, such a behavior of ACF (ATP-utilizing Chromatin remodeling and assembly Factor), a human variant of ISWI, has been observed by monitoring the sliding of DNA on the histone with FRET (Fluorescence Resonance Energy Transfer) in \emph{in vitro} preparations  \cite{Yang2006,Racki2009,Hwang2014}. A basic requirement for this molecular function is, therefore, their ability to detect the lengths of the DNA chains that flank a nucleosome. 
 
These \emph{in vitro} experiments show the motor's range of sensitivity to chain length. Initially a nucleosome is placed at one end of DNA by sequence affinity \cite{arneodo2011}. The nucleosome starts to translocate toward the middle of DNA chain stepwise after a brief period of several seconds if the flanking DNA on the other side is longer than $\sim$20 bp on addition of ATP \cite{Yang2006,Leonard2015}. The speed of translocation depends on the length of the flanking DNA and saturates where the length is $\sim$60 bp\cite{Yang2006}, corresponding to the saturation of the exit rate from the ``pause'' phase of the stepwise translation, observed with high time resolution \cite{Hwang2014}. This length approximately corresponds to the persistence length of DNA.  

Until now very few physical models  have been proposed for the action of chromatin remodeling factors \cite{Chou2007, Schiessel2009, Chowdhury2012,Blossey2012,Blossey2013}, probably due to insufficient experimental data on the same remodeler species. Indeed, it is not yet clear whether or not different versions of the same family of remodelers have the same functioning mechanisms.  Most of these works are either generic models, describing how an ATP-consuming complex can displace a nucleosome, and in the case of ACF,  length-dependent reaction rates are assumed \cite{Blossey2012, Blossey2013}. 

How can the length of flanking DNA be detected? ACF is composed of an ATPase domain and several flanking domains, which are specific to each remodeler \cite{Clapier2009}. The motor monomer has a domain, which binds to extranucleosomal DNA. We can exclude a mechanism, where the motor makes contact with the end of the DNA strand because the persistence length of DNA is about 50 to 60 nm, larger than the dimension ($R\sim 8$ nm) of the motor. Thus, it must detect the chain length by  some local interaction with DNA. Here we hypothesize that ACF's length sensitivity is  due to the binding of the motor to the non-nucleosomal DNA, which fluctuates around the nucleosome. 

Since the length of DNA that we need to describe is up to its persistence length, we assume the DNA chains can be treated as a beam, described by Euler-Bernoulli theory. The motors are very large dimers with molecular mass of $\sim$1 MDa, covering the surface of histone octamers \cite{Leschziner2011}. The subunits of a dimer may interact with each other \cite{Racki2009} in a certain way not to waste energy by playing a ``tug-of-war'' between them by acting independently \cite{Leonard2015}. We first focus on the way how a single subunit can detect the length of DNA and then come back to the issue as to how two subunits could interact with each other.

\section{FLUCTUATIONS OF DNA}
The binding of DNA to a histone octamer is not static but undergoes thermal fluctuations called ``breathing,'' which uncovers DNA located at the edge of the octamer at a certain probability \cite{Li2005}. The rate of unwrapping is $\sim$4/s and that of re-wrapping is $\sim 50$/s for the system with extremely short flanking chains \cite{Li2005}. Given the stiffness of DNA chain, this mode of motion can be described as a pivoting motion of the DNA chain if the chain length is less than the persistence length of about 50 nm (Fig.\ \ref{fig:fluc}B). 

The remodeling motor that is associated with the histone octamer is much larger than the octamer and has DNA binding sites. We could assume that binding and unbinding of DNA to the motor's binding site are related to wrapping and unwrapping of DNA on a histone octamer. However, the correspondence may not be one-to-one. The observed higher rate for unwrapping is likely localized, involving on average a shorter distance for the chain to travel than wrapping. Since the binding site is likely some distance away from the histone surface, unbinding may requite a larger travel than unwrapping, the unbinding rate could be lower than the unwrapping rate. Thus the binding rate and the unbinding rate are likely closer to each other than the winding rate and the unwinding rates are. To describe the motion of the chain associated binding and unbinding in a simple manner, 
it would be useful to assume a quadratic potential well with respect to the relative orientation of the chain formed by a holding spring with relatively shallow local minima at two positions off center. If these local minima are shallower than the thermal energy, pivoting motion of the polymer can be described by the stiffness of the holding spring, which in turn determines the transition rates between two states, one of which is ``bound'' to the receptor site of the motor molecule and the other ``unbound.''  

One of these states can be the starting point of an ATP-dependent sliding motion between the DNA and nucleosome. The dependence on the chain length is introduced through the viscous drag on the chain fluctuations during transitions between ``bound'' and ``unbound'' states, which require movement of the chain assuming that the nucleosome is stationary. 

If we assume that the DNA chain, which is approximated by an Euler-Bernoulli beam, is held by a pivot with a spring at the surface of the nucleosome, the dominant mode of fluctuations is pivoting motion and the amplitude of angular fluctuations shows little length dependence. Let $\theta$ the angular amplitude of the polymer around the mean position. The amplitude of the dominant fluctuation is determined by the spring $k_\theta$ that holds DNA,
\begin{equation}
\frac 1 2 k_\theta\langle \Delta\theta^2\rangle\sim k_BT,
\end{equation}
and does not depend on the chain length $L$.
If the rotational drag of the nucleosome is smaller than that of the DNA, those transitions take place mainly by a rotation of the nucleosome, insensitive to the chain length of DNA.

The pivoting motion is described by,
\begin{equation}
\zeta_r \frac{d\theta(t)}{dt}=-k_\theta \theta(t)+F(t),
\end{equation}
where $F(t)$ is random force and $\zeta_r$ is the rotational diffusion coefficient of the polymer chain. For ``white'' random noise, the power spectrum $S_\theta(\omega)$ is,
\begin{equation}
S_\theta(\omega)=\frac{2k_BT}{\zeta_r}\cdot\frac{1}{(k_\theta/\zeta_r)^2+\omega^2},
\end{equation}
where $\omega$ is angular frequency. The spectrum has a characteristic frequency $k_\theta/\zeta_r$.

Rotational diffusion coefficient $D_r$ of short DNA chains is well approximated by assuming short DNA chains as a rod \cite{tirado1984}. The analytical expression is,
\begin{eqnarray} 
\pi\eta L^3D_r/(3k_BT)=\ln(L/2r)+\delta,
\end{eqnarray}
where $\delta\approx -0.7+O(2r/L)$. The rotational drag coefficient $\zeta_r$ of a DNA chain is thus,
\begin{eqnarray}
 \zeta_r\approx\frac 1 3\pi\eta L^3/(\ln(L/2r)-0.7),
 \label{eq:zeta}
\end{eqnarray}

\noindent using Einstein's relationship. The terms $O(2r/L)$ could be ignored if $L\gg2r$. For DNA, $2r\approx 2$nm.

For the chain length of the DNA to be important for anglular fluctuations at the supporting point,  the rotational motion of the nucleosome must be less than the chain fluctuations of the DNA, namely the rotational drag of the nucleosome must be larger than that of DNA chain. We show later that this factor can impose a detection limit.

\section{THE MODEL}

\begin{figure}[h] 
\begin{center}
\includegraphics[width=7cm]{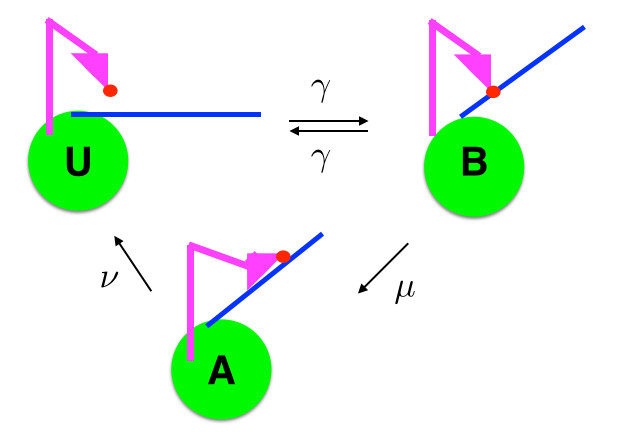}
\caption{A schematic illustration of the three states of the monomer: \textbf{U}nbound, \textbf{B}ound, and \textbf{A}dvanced. The motor attached to the histone octomer (circle) extends its arm (bent line with a triangle) toward the DNA chain (straight line), which undergoes pivoting motion driven by thermal energy. The rate of transition between the bound state and the unbound state is $\gamma$ in both directions.}
\label{fig:states}
\end{center}
\end{figure}

We have already discussed two states, ``unbound'' and ``bound'' (Fig.\ \ref{fig:states}). For describing a motor, we need to provide another state that describes motile activity. The proposed model consists of the following assumptions.

\subsection{The assumptions}
\begin{enumerate}
\item The binding site of the motor has three states, bound (B), unbound (U), and advanced (A).
\item Transitions between B and U depends on the contact of DNA to the site and thus depends on the movement of DNA.
\item Transitions from B to A and from A to U involves ATP binding and hydrolysis. For this reason these transitions are unidirectional.
\item Transition from U to A produces unidirectional movement of DNA relative to the nucleosome.
\end{enumerate}

\subsection{The equations}
The transitions between these three states are represented by $\gamma$, $\mu$, and $\nu$ (Fig.\ \ref{fig:states}).
The transition rate $\gamma$ between U and B in both directions  is expressed by $\gamma \sim k_\theta/\zeta_r$, reflecting thermal fluctuations of the DNA. Since $\zeta_r$ is an increasing function of the chain length $L$ (Eq.\ \ref{eq:zeta}),  the rate $\gamma$ is a decreasing function of $L$. Other rates $\mu$ and $\nu$ are independent of $L$. The differential equations that govern the transitions between these states are,

\begin{eqnarray}\label{eq:deqU}
\frac{dU}{dt}&=& -\gamma U+\gamma B + \nu A,\\ \label{eq:deqB}
\frac{dB}{dt}&=&\; \gamma U -(\mu+\gamma) B, \\ \label{eq:deqA}
\frac{dA}{dt}&=& \qquad\quad\; \mu B- \nu A,
\end{eqnarray}

\noindent where $\mu$ and $\nu$ are transitions, in which ATP binding and its hydrolysis are involved. The rotation rate of the motor is proportional to $\nu A$ for normalized A, i.e. $U+B+A=1$.

\section{RESULTS}
Here we present the solution to the problem and then briefly state its implication.
\subsection{The solution}
The Jacobian of Eqs.\ \ref{eq:deqU}--\ref{eq:deqA} has eigenvalues $0$ and $ 1/ 2 -(2\gamma-\mu-\nu)\pm\sqrt{4\gamma^2+(\mu-\nu)^2-4\gamma\nu}$.
The eigenvalues other than $0$ correspond to transient modes. The eigenvector for eigenvalue $0$ is then given by 
\begin{equation*}
\left(
\begin{array}{c} U\\ B\\ A
\end{array}
\right)=\left(
\begin{array}{c} \gamma+\mu\\ \gamma\\ \gamma\mu/\nu
\end{array}
\right).
\end{equation*}

\noindent After normalization of this vector, we obtain for the rotation rate of the motor,
\begin{eqnarray}
\nu A=\frac{\gamma\mu}{\sqrt{\gamma^2+(\gamma+\mu)^2+(\gamma\mu/\nu)^2}}.
\label{eq:nuA}
\end{eqnarray}

\begin{figure}[h]
\centering
\includegraphics[width=7cm]{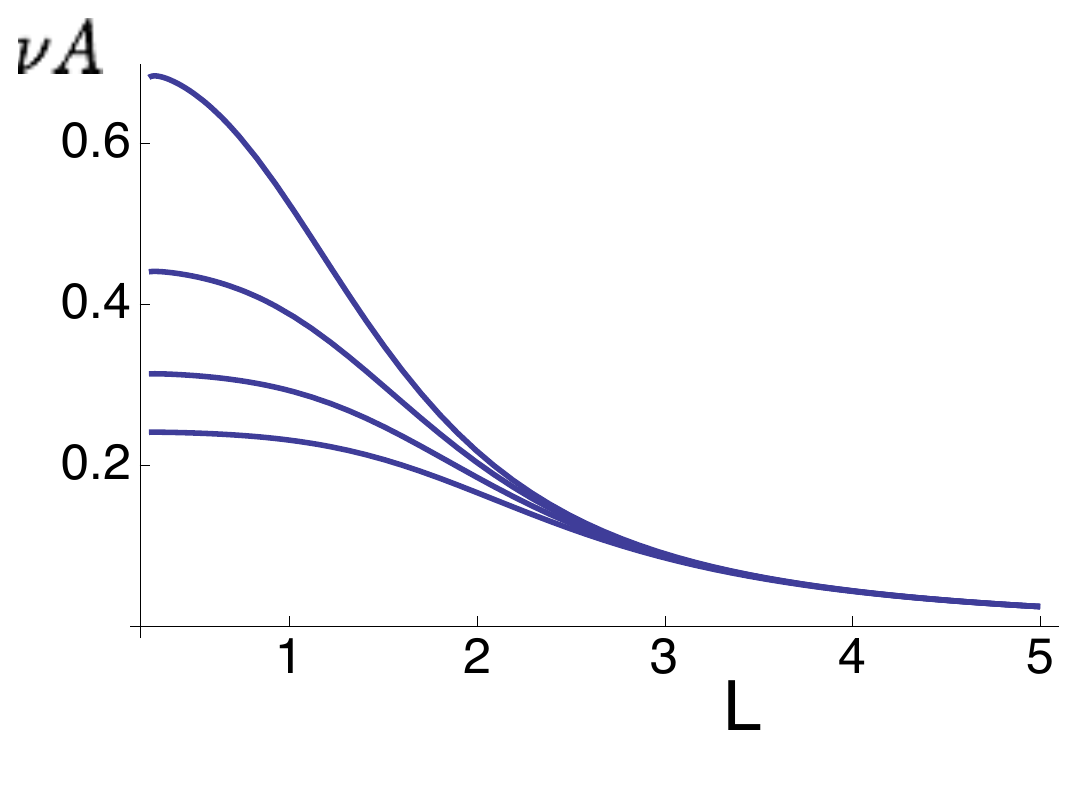}
\caption{The advance speed $\nu A$ of the motor  plotted against the chain length $L$ for $\mu=1$. Traces are from the top, $\nu\rightarrow\infty$, $\nu=1, \nu=1/\sqrt 2,\; \mathrm{and}\; \nu=1/2$. Notice that $L$ in the plot is a parameter that is proportional to the chain length because $\gamma \sim (\ln (L/2r)-0.7)/L^{3}$ was used instead of $\gamma\sim k_\theta/\zeta_r$ to show the functional dependence on $L$. For the scale of the abscissa, see the subsection on transition rates in the text. The value of 0.1 is assumed for $r$.}
\label{fig:L_dep}
\end{figure}

The last term of the denominator diminishes for large $\nu$. If $u$ is the unit distance the motor travels per cycle, the speed of the motor, which is expressed as $\nu Au$, is an increasing function of $\gamma$ expect for large values of $\nu$, where it behaves as a constant. 

If we assume $\gamma \sim (\ln (L/2r)-0.7)/L^{3}$, the advance speed of the motor is inversely related to the chain length  (Fig.\ \ref{fig:L_dep}). This relationship provides a basis for length detection by a motor monomer.

\subsection{The implication}
How this outcome of our model is related to experimental observations? Suppose hypothetically that a single remodeling motor is attached to a nucleosome and the nucleosome is at one end of DNA. If the DNA is long enough, the motor pushes away the DNA. As the DNA becomes longer, the speed is reduced, as shown in Fig. \ref{fig:L_dep}. However, the motor functions as a dimer.  Assume that a monomer is located on the right of the nucleosome and the other on the left of it. 

Let $L_\ell$ be the length of DNA flanking on the left of the nucleosome and the length of the one flanking on the right  $L_r$. If the motors act independently of each other, the leftward speed $(\nu A)_\ell$ would be determined by $L_\ell$, and the rightward speed $(\nu A)_r$ by $L_r$, as described by the functional dependence plotted in Fig. 3. However, the two subunits compete with each other to move a single DNA in opposite directions. Thus, the probability of moving the nucleosome to the left would be an increasing function of $(\nu A)_\ell-(\nu A)_r$. That, in turn, is an increasing function of $L_r-L_\ell$ because $\nu A$ is a decreasing function of $L$. If $L_r>L_\ell$ the nucleosome moves to the right and the speed increases with the difference $L_r-L_\ell$. Thus, not only the direction but the speed is determined by the length difference, consistent with experimental observations \cite{Yang2006,Hwang2014}. 

If the subunits work independently of each other as discussed above, the competition between them takes place after ATP hydrolysis, chemical energy is wasted in a ``tug-of-war.'' However, the subunits can compete with each other before the ATP hydrolysis step, such energy loss can be prevented. An example of such dimerization is discussed later.

In some in vitro experimental configurations, a nucleosome is initially located at one end of DNA ~\cite{Yang2006,Hwang2014}, and it may appear as if no chain is on one side. It would be reasonable to regard such a situation as a limiting case for short chain length and to expect continuity from short chains. 

We notice that the values for the transition rates must be realistic for our model to be meaningful. For example, the model does not lead to a meaningful length sensitivity if fluctuations are very fast, making $\gamma$ much larger than $\mu$. The same thing happens if the transition rate $\nu$ is small compared with other rates.  This issue will be addressed in Discussion.

\section{DISCUSSION}
We sought the source of chain length dependence to pivoting fluctuations of DNA in our model. There is an earlier report that suggests that pivoting fluctuations of rigid polymers are associated with receptor binding in length dependent manner \cite{Kalinina2013}. In that system, a longer spindle microtubule binds to a kinetochore faster than shorter ones because binding sites are distributed uniformly in space, giving a longer polymer, which sweeps a larger volume in a given time, a larger chance of encountering a binding site \cite{Kalinina2013}. In contrast, binding sites in our system are highly localized near the pivoting point, making the frequency of fluctuations the key factor for binding.

\subsection{Transition rates}
As mentioned earlier, a critical test of the present model is whether or not we can assign realistic values for transition rates. We have assumed that the fluctuation rate $\gamma$ is somewhat lower than ``breathing,'' thermal fluctuations of DNA at the surface of a histone octamer \cite{Bintu2012,Hwang2014,Ngo2015}. The experimental values are $\sim4$ s$^{-1}$ for unwinding and $\sim50$ s$^{-1}$ for rewinding, using DNA with 147 bp \cite{Li2005}, just long enough to wrap around a histone octamer. 
Since we assumed symmetric rates 
on the ground that the binding site is located some distance away from the histone octamer, 
it would be reasonable to expect $\gamma \sim$4 s$^{-1}$ because the lower rate in the wrapping-unwrapping transitions should be rate limiting for binding-unbinding transitions.  The time required for completing a single cycle of the motor is $\sim$10 s \cite{Hwang2014}. These experimental values can be compatible with the value of $\sim$1 s$^{-1}$ for $\mu$. In addition, the motile cycle was measured for dimers and competing monomers could slow down the motile activity. The value of $\nu$ would be larger than $\mu$. This examination indicates that realistic parameter values can be consistent with observed length dependence of the motor.

In a series of experiments with high time resolution, the movement of the nucleosome is stepwise, consisting of two phases, a stationary phase and a translocating phase \cite{Hwang2014}. Length dependence is observed only in the duration of the stationary phase. The duration of the translocating phase is relatively short ($\sim 2$s) and the speed of translocation is constant ($\sim15$ bs/s). Neither the duration nor the speed shows chain length dependence \cite{Hwang2014}. In our model, the stationary phase corresponds to the duration of the motor staying in the states U and B. This duration is longer for larger chain length $L$ (with smaller $\gamma$) because for a smaller $\gamma$ the probability of the motor in state B is less, leading to a lower transition rate to A. On exiting from state B to A, the motor translocates the nucleosome. This process is independent of $L$ because it does not depend on $\gamma$ for a large $\nu$. Thus, our model is compatible with these experimental observations \cite{Hwang2014}. 

We assumed symmetric rates for binding and unbinding in the treatment. That is, however, not a critical assumption that leads to the length dependence derived. Qualitatively the same chain length dependence can be derived even if only the binding rate is chain length dependent. 

\subsection{Stiffness of motor arm and DNA}
The schematic illustration (Fig.\ \ref{fig:states}) might give an impression as if we assumed that the receptor arm of the motor is stiffer than DNA. That is, however, not a necessary condition that leads to our conclusion. The necessary condition is that the receptor of the motor has a localized distribution, which is asymmetric with respect to the mean position of the DNA chain. 

The predicted direction of the DNA away from the motor (Fig.\ \ref{fig:states}) in the model could be interpreted that a compressive force needs to be applied to the arm of the motor, requiring considerable stiffness. However, the force for sliding the DNA could be applied at other sites of the motor after the receptor binding by allosteric interaction. 

\subsection{Rotational diffusion of the nucleosome}
Pivoting motion of the polymer chain is meaningful only if the nucleosome is large enough so that the nucleosome does not rotate together with the polymer chain. An order estimate of the rotational diffusion constant of the nucleosome can be obtained by assuming spherical geometry with the radius $R$ that is consistent with the molecular mass and the density of water. An estimate of $R\approx 8$nm is obtained from the molecular mass of 1 MDa, most of which comes from the motor itself.  A histone octamer has molecular mass of 108.8 kD, consisting of two copies of each core histone proteins, H2A (14.3 kD), H2B (13.8 kD), H3 (15 kD), and H4 (11.3 kD). The rotational diffusion constant for the sphere $\zeta_n=8\pi\eta R^3$ leads to, the ratio of the diffusion constants $\zeta_n/\zeta_r=24(R/L)^3[\ln(L/2r)-0.7]$.

The ratio $\zeta_n/\zeta_r$ is approximately 0.3 for $L=40$ nm. However, it is likely that the value used for $\zeta_n$ is an underestimate because the nucleosome is not spherical and its surface is unlikely smooth. For this reason, the ratio would be favorable enough for $L\approx40$ nm. However, the ratio may impose a detection limit for smaller chain length.

\subsection{The range of length sensitivity}
The motor does not move if the nucleosome is located at the end of DNA and the length of flanking DNA is less than 20 bp. There are two possible reasons.
One is that the slope of $\nu A$ is small at small $L$ (Fig.\ \ref{fig:L_dep}). The small slope makes the difference in $\nu A$ in two sides too small to detect. Another reason for the lower detection limit is that the pivoting mode of chain fluctuations may not be larger than rotational diffusion of the nucleosome. 

The upper limit of $\sim 60$ bp can be attributed to the persistence length of DNA. Fig.\ \ref{fig:L_dep} does show saturating behavior for long chains. However, it is hard to determine the saturating physical chain length because $\gamma\sim k_\theta/\zeta_r$ not only depends on $\zeta_r$, which is determined by $L$ alone but also $k_\theta$, which needs to be determined experimentally.

Sequence specificity of DNA could affect the activity of the motor because such a specificity, which creates a well in the potential energy \cite{arneodo2011}, was used for the initial placement of the nucleosome at an end of DNA for the \emph{in vitro} experiments \cite{Yang2006,Racki2009,Hwang2014}. Such a sequence specificity has not been observed during \emph{in vitro} experiments \cite{Yang2006,Partensky2009}. Nonetheless, it could be involved in some repositioning of nucleosomes while the concentration of ATP is low.

\begin{figure}[h]
\begin{center}
\includegraphics[width=7cm]{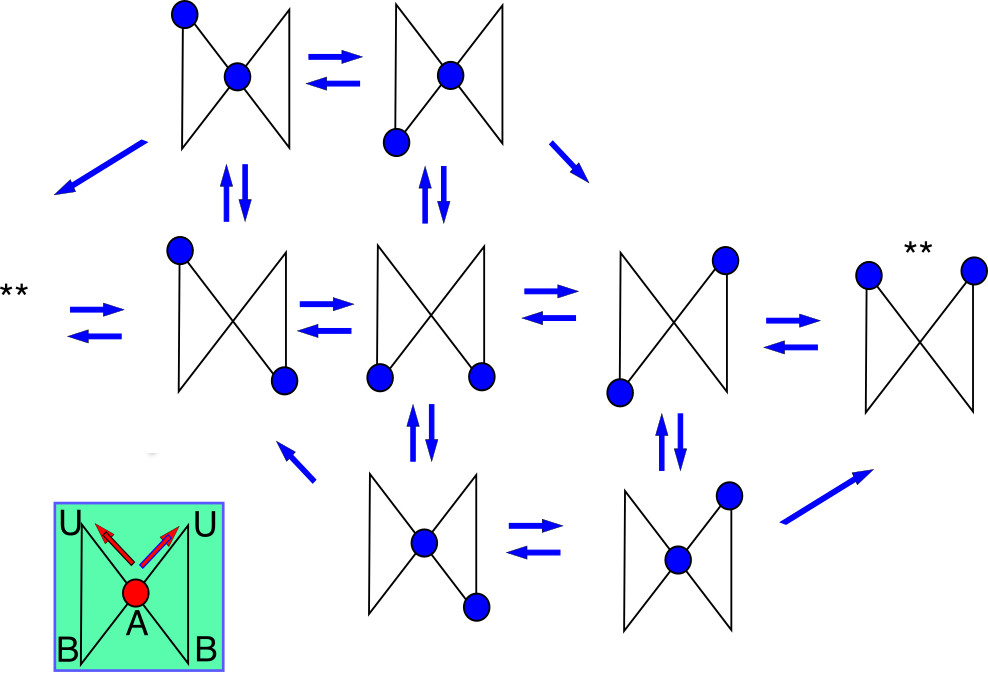}
\caption{An example of dimerization.  Two monomers are combined symmetrically as indicated in the box. Filled circles indicate the state of each monomer, either U, B, or A, represented by a triangle. See the box on the lower left. Only one of the monomer can be in state A to prevent ``tugs-of-war.'' A single state marked with ** appears in two places. Arrows indicate direction of transitions.}
\label{fig:dimer}
\end{center}
\end{figure}

\subsection{Dimerization}
So far we have described the monomer, while the remodeling motor consists of a dimer. If each monomer works independently of each other, the efficiency of the motor would be very low because of a tug-of-war should take place. It has been suggested recently that such a problem can be avoided by a shared binding site between them \cite{Leonard2015}. This arrangement can be easily implemented in our model by combining the two subunits symmetrically with a shared state, resulting in an 8-state model (Fig.\ \ref{fig:dimer}).

The present model is aimed at explaining \emph{in vitro} experiments not the \emph{in vivo} mechanism. However, it is tempting to speculate that the presented mechanism could have some relevance to \emph{in vivo} function in view of the recent findings of linear aggregations of nucleosomes \cite{Fudenberg2012,Ricci2015}. 
Such aggregations could allow fluctuations of free DNA chains between the nucleosomes, even though what fluctuate are loops instead of single chains as described here. If the fluctuations of loops can be treated similarly to those of single chains, then the remodeling motor could pull back long loops of DNA, equalizing the length of all the loops.

\section{CONCLUSIONS}
The present paper presents fluctuations of DNA interacting with a DNA binding site of a protein as a possible mechanism for ACF, a member of the ISWI family of nucleosome remodelers, to detect the lengths of the DNA chains flanking the nucleosome. While this motor functions as a dimer, the model presented here is basically that of a monomer.  The immediate next task is to extend the present description to a dimer, which is given only a brief description here. Another task would be the incorporation of the sensing mechanism presented here into a physical model, which describes motile activity.

From experimentalist perspective, the present model provides a number of testable predictions. The chain length dependences of ``breathing'' and that of binding-unbinding to the motor molecule would be a set of such testable predictions. The most critical prediction of the model is, however, that the motor binds to the shorter side of flanking DNA before translocation. Predictions of detailed steps in the translocation process may become available only after dimerizing the present monomer model. 

\section*{Acknowledgments}
The most part of this work presented was done at Max Planck Institute for Physics of Complex Systems. We thank Dr.\ Frank J\"{u}licher for his hospitality and stimulating discussion. AMF thanks Prof.\ Geeta  Narlikar for a short stay in her lab and Dr.\ Angelo Rosa for critical reading of the manuscript. Thoughtful comments by the reviewers are greatly appreciated.

\section*{References}
\bibliographystyle{iopart-num}
\providecommand{\newblock}{}

\end{document}